\documentclass[twocolumn,showpacs,prl,10pt]{revtex4}
\usepackage[english]{babel}
\usepackage{amsmath,amssymb}
\usepackage{times}
\usepackage{epsfig}

\begin{document}

\title {Freezing of spin dynamics in underdoped cuprates}
\author{I. Sega$^1$ and P. Prelov\v sek$^{1,2}$}
\affiliation{$^1$J.\ Stefan Institute, SI-1000 Ljubljana,
Slovenia}
\affiliation{$^2$Faculty of Mathematics and Physics, University
of Ljubljana, SI-1000 Ljubljana, Slovenia} 
\date{\today}

\begin{abstract}
The Mori's memory function approach to spin dynamics in doped
antiferromagnetic insulator combined with the assumption of
temperature independent static spin correlations and constant
collective mode damping leads to $\omega/T$ scaling in a broad
range. The theory involving a nonuniversal scaling parameter is used
to analyze recent inelastic neutron scattering results for underdoped 
cuprates. Adopting modified damping function also the emerging central peak 
in low-doped cuprates at low temperatures can be explained within the same
framework. 
\end{abstract}

\pacs{71.27.+a, 75.20.-g, 74.72.-h}
\maketitle

It is experimentally well established that magnetic static and
dynamical properties of high-$T_c$ cuprates are quite anomalous. Even
at intermediate doping the magnetic response does not follow the
Fermi-liquid phenomenology in the metallic state, and reveals the
remarkable resonant peak within the superconducting (SC) state.  With
decreasing doping the magnetic response approaches in a novel way the one 
of the reference undoped material--the  antiferromagnetic
(AFM) insulator. It has been first observed by inelastic neutron
scattering (INS) experiments on low-doped La$_{2-x}$Sr$_x$CuO$_4$
(LSCO) \cite{keim,kast} that local ({\bf q}-integrated) spin dynamics
in the normal state exhibits anomalous  $\omega/T$ scaling, not reflected in the AFM correlation 
length $\xi$ which shows no significant $T$-dependence.
Similar  behavior has been found also in YBaCu$_3$O$_{6+x}$ (YBCO), initially 
at $x=0.5, 0.6$ \cite{ster} with the regime restricted to $T>T_c$, and in
Zn-doped YBCO \cite{kaku} with reduced $T_c \sim 0$. 

Recent, more detailed INS experiments on heavily underdoped (UD),
including Li-doped LSCO \cite{bao}, YBCO with $x=0.35$
\cite{stoc1,stoc2} and $x=0.45$ \cite{hink1,hink2}, as well as
electron-doped Pr$_{1-x}$LaCe$_x$CuO$_{4-\delta}$ (PLCCO) with
$x=0.12$ \cite{wils1,wils2}, confirm universal features of anomalous
normal-state spin dynamics. Latter materials exhibit no SC (Li-doped
LSCO) or have very low $T_c$ (YBCO, PLCCO) so that the $\omega/T$
scaling is found in a broad range both in ${\bf q}$-integrated local
susceptibility $\chi''_L(\omega)$ \cite{bao,stoc2,hink2} and in
$\chi''_{\bf q}(\omega)$ at the commensurate AFM ${\bf q}={\bf Q}=(\pi,\pi)$
\cite{bao,stoc2,wils1,wils2}. Typically, $\chi''_{\bf Q}(\omega)$ is a
Lorentzian with the characteristic relaxation rate scaling as $\Gamma =
\alpha T$, but with a nonuniversal $\alpha$. Similarly, normalized
$\chi''_L(\omega)$ has been fitted to the scaling form 
$f(\omega,T)=\chi''_L(\omega,T)/\chi''_L(\omega,T=0) \sim (2/\pi) {\rm
tan^{-1}}(A_1 \omega/T+A_2(\omega/T)^3)$ \cite{stoc2,hink2} with material
dependent $A_{1,2}$. It is characteristic that the normal-state-like scaling is found
even below $T_c$ without any clear sign of the presence of the resonant
peak. On the other hand, it has been observed that at low $T$ the spin
dynamics gradually transfers into a quasi-elastic peak for $T<T_g$
\cite{bao,stoc1} whereas the inelastic response saturates. However, this
{\it freezing} mechanism appears to be of entirely dynamical origin since
$\xi$ as well as the integrated intensity are unaffected by the crossover. 

In order to describe the anomalous scaling of spin dynamics as well as
of other electronic properties of cuprates the concept of marginal
Fermi liquid has been introduced \cite{varm}. Whereas the deeper origin
of such behavior  is often given in terms of proximity to a QCP
\cite{chak,aepp} the above systems clearly lack the criticality of
$\xi(T)$. It is thus quite remarkable that a similar behavior is also found in
heavy-fermion metal system CeCuAu \cite{schr} which possesses a well defined
QCP (unlike in cuprates) in the phase diagram, nevertheless exhibits similar 
$\omega/T$ scaling without the criticality in $\xi(T)$.

The present authors introduced a theory of spin dynamics in doped AFM
\cite{prel} which describes the scaling behavior as a dynamical
phenomenon. Namely, assuming that static quantities--the equal-time spin
correlations in particular--are unaffected (or weakly
dependent) by $T$, and that the system is metallic with finite spin
collective-mode damping, the system close to AFM naturally exhibits
$\omega/T$ scaling in a wide energy range, but unlike at usual QCP,
with a saturation at low-enough $T$. It should be noted that such
a scenario is close to the theory of freezing of a liquid, which  is
as well a dynamical phenomenon reflected only weakly in equal-time
correlations.

The aim of this paper is to analyse recent INS experiments on
different cuprates mentioned above within this framework and show that
observed differences (nonuniversality) in scaling behavior can be made
consistent taking into account actual parameter regimes. Moreover, the
analysis can be generalized to the emergence of the CP assuming same static
quantities but the change in collective-mode damping. 

Within the memory-function approach
\cite{prel,sega} the dynamical spin susceptibilty can be
generally expressed as
\begin{equation}
\chi_{\bf q}(\omega)=\frac{-\eta_{\bf q}}{\omega^2+\omega 
M_{\bf q}(\omega) - \omega^2_{\bf q}}\,. 
\label{chiq}
\end{equation}
The spin stiffness $\eta_{\bf q}$ is the first freqency moment of the dynamical susceptibility
$\eta_{\bf q}\!=\!-{\dot\iota}\langle [S^z_{-\bf q}\, ,\dot{S}^z_{\bf q})]\rangle$,
and can be within any relevant model expressed in terms of equal-time
correlations only, whereas $\omega_{\bf q}=(\eta_{\bf q}/\chi^0_{\bf q})^{1/2}$ is an 
effective collective mode frequency where $\chi^0_{\bf q}\!=\chi_{\bf q}(\omega\!=0)$ is 
the static susceptibility.Since we have in mind a doped AFM the actual
evaluation in relevant models like the $t$-$J$ model \cite{prel,sega} reveal
that $\eta_{\bf q}$ is only weakly ${\bf q}$-dependent for ${\bf q} \sim {\bf Q}$
whereby $\eta_{\bf q}\sim\eta \sim 2J$.

The form (\ref{chiq}) is particularly suited for a description of 
damped collective modes close to an ordered magnetic state. $M_{\bf q}(\omega)$ is the 
(complex) memory function incorporatinginformation on the collective-mode
damping $\gamma_{\bf q}(\omega)=M^{\prime\prime}_{\bf q}(\omega)$. Since we will be 
here dealing withdoped cuprates in the normal state which are (anomalous)
paramagnetic metals (PM), low-frequency collective modes at ${\bf q} \sim {\bf Q}$ are 
generally overdamped $\gamma_{\bf q}>\omega_{\bf q}$ consistent with INS results.

Underdoped cuprates close to the AFM phase represent a system of low 
charge-carrier concentration but large spin fluctuations whose dynamics is
quite generally restricted by the sum rule
\begin{equation}
\frac{1}{\pi}\int_0^\infty d\omega ~{\rm cth}\frac{\omega}{2T}
\chi^{\prime\prime}_{\bf q}(\omega)= 
\langle S^z_{-{\bf q}} S^z_{\bf q}\rangle = C_{\bf q}\, .
\label{eqsum}
\end{equation}
In doped AFM $C_{\bf q}$ is strongly peaked at ${\bf Q}$ with
a characteristic width $\kappa_T=1/\xi$. Moreover, the total sum rule is
for a system with local magnetic moments (spin $1/2$) given by
$(1/N)\sum_{\bf q} C_{\bf q} = (1-c_h)/4$, where $c_h$ is an effective hole
doping \cite{sega}.  

While the formalism so far is very general, we now introduce
approximations specific to UD \cite{prel}. INS experiments listed above
indicate that within the normal state the effective ${\bf q}$ width of
$\chi_{\bf q}''(\omega)$, i.e., dynamical $\kappa(\omega)$, is only weakly
$T$- and $\omega$-dependent, even on entering the regime with the CP response which
implies that $C_{\bf q}$ is $T$ independent or at least not critical
for $T \to 0$. This  has been also confirmed be the present authors \cite{prel}
in numerical investigations of the relevant $t$-$J$ model, where $\kappa_T$
has been found to be rather large even for the lowest doping. 
Since $\eta_{\bf q}\sim \eta$ is also nearly constant  the $\omega/T$
scaling and the freezing mechanism must emerge from spin
dynamics not reflected in equal-time correlations.  
Within further analysis we will assume the commensurate AFM response at
${\bf Q}$ and the double-Lorentzian form $C_{\bf q}=C/[({\bf
q}-{\bf Q})^2+\kappa_T^2]^2$ close to INS experiments although qualitative
results at low $\omega$ do not depend on a particular form of $C_{\bf q}$.

\noindent {\it Paramagnetic metal:} Even at rather low doping cuprates
behave as PM, as manifest, e.g.,  by their metallic
resistivity $\rho(T)$ and quasiparticle-like excitations on (the parts
of) the  Fermi surface as revealed by ARPES. This is underlying our
assumption that the (Landau) damping $\gamma_{\bf q}(\omega)$ is
dominated by particle-hole excitations \cite{sega} being weakly ${\bf q}$
and $\omega$ dependent, hence $\gamma_{\bf q}(\omega) \sim \gamma$, as has
been found also in the numerical analysis of model 
results \cite{prel}, whereby a noncritical dependence $\gamma(T)$ can
still play some role. We treat here $\gamma$ as a phenomenological
parameter although it can be estimated from model
results \cite{prel} or even analytically \cite{sega}. It could be also
extracted from high-energy INS experiments following 
the crossover to underdamped modes $\omega_{\bf q}\sim\gamma$. However, most
INS data on UD cuprates  are restricted to energies $\omega<50$~meV being too low
to observe the crossover.  

The assumption of constant $\gamma$ in Eq.~(\ref{chiq}) leads to 
\begin{equation}
\chi''_{\bf q}(\omega)=\frac{\eta \gamma \omega}{(\omega^2 -
\omega^2_{\bf q})^2 + \gamma^2 \omega^2}\,, 
\label{dosc}
\end{equation}
which in the overdamped regime, $\omega_{\bf q}<\gamma/2$ and
$\omega\ll\gamma$, results in   
\begin{equation}
\chi''_{\bf q}(\omega) \sim \chi_{\bf q}^0\frac{\omega\Gamma_{\bf q}}
{(\omega^2 + \Gamma^2_{\bf q}) } , \quad \Gamma_{\bf q}=
\frac{\eta}{\gamma\chi^0_{\bf q}}.
\label{chiim1}
\end{equation}
While such form appears plausible, it can
be explicitly tested experimentally.  Recent INS data obtained for
$\chi''_{\bf Q}(\omega)$ have been fitted with
Eq.(\ref{chiim1}) \cite{bao,stoc2} to extract $\Gamma_{\bf Q}(T)$. Indeed,
plotting the interdependence of data for $\chi^0_{\bf Q}$ and $\Gamma_{\bf Q}$
reported in \cite{stoc2} (see Fig.~12) shows that $\chi^0_{\bf Q}\Gamma_{\bf
Q}(T)={\rm const.}(=\eta/\gamma)$ to within experimental error over the {\it
whole} span of $\Gamma_{\bf Q}$, confirming in this way our assumption of
constant $\gamma(\omega) \sim \gamma$.

Let us now  concentrate on the behavior of $\chi''_{\bf  Q}(\omega)$ as
enforced by the sum-rule, Eq.(\ref{eqsum}). It has been noticed that the variation
of $\Gamma_{\bf Q}$ is mainly determined by the parameter 
\begin{equation}
\zeta=C_{\bf Q}\pi\gamma/(2 \eta), \label{zeta}
\end{equation}
and only marginally on, e.g., $\gamma$ provided that we are discussing
the regime $T<\!<\gamma$ which is experimentally relevant.
\begin{figure}[h!]
\centering
\epsfig{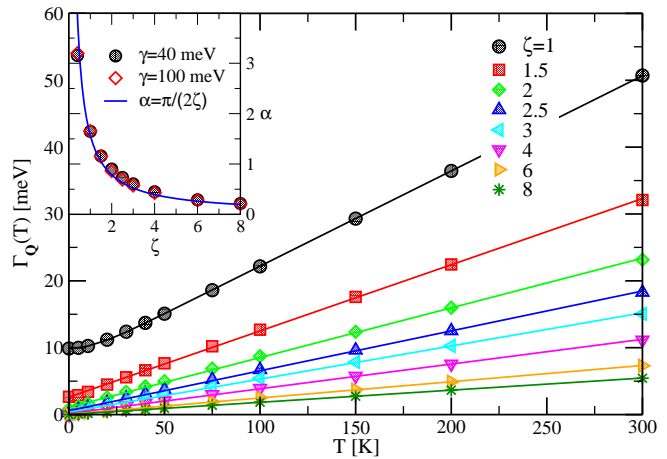}
\caption{Relaxation rate $\Gamma_{\bf Q}$ vs. $T$ for different 
parameters $\zeta$ and $\gamma$=40 meV. Inset: dependence of slope parameter
$\alpha$ on $\zeta$ for two different $\gamma$. For comparison, an estimate
for $\alpha$ based on Eq.~(\ref{zeta1}) is included.}
\label{fig1}
\end{figure}

In Fig.~1 we first present results for $\Gamma_{\bf Q}(T)$ in the
experimentally relevant regime and units for a range of $\zeta=1 -
8$.  It is evident that for any $\zeta$ there exists a finite limiting value
$\Gamma^0_{\bf Q}\!=\!\Gamma_{\bf Q}(T\!\to\!0)$ being strongly dependent on
$\zeta$, i.e., $\Gamma^0_{\bf Q} \sim \gamma {\rm exp}(-2\zeta)$ \cite{prel}.  On
the other hand, for $T> \Gamma^0_{\bf Q}$ the variation is nearly linear
$\Gamma_{\bf Q} \sim \alpha T$ being a manifestation of the $\omega/T$
scaling. It is also seen that the linear-in-$T$ dependence of $\Gamma_{\bf
Q}$ extends well beyond the limit $\omega\sim\gamma/2$ provided, of course,
that $\kappa_T$ remains weakly dependent on $T$ in the respective temperature
region. 

As seen in Fig.~1, the slope $\alpha$ is not universal but depends on
$\zeta$ as well as on $\gamma$, where the latter dependence is very weak
(logarithmic) as is also manifest from the collapse of the data for two values of
$\gamma$ onto a single set (inset to Fig.~1). In fact, from
Eqs.~(\ref{eqsum}) and (\ref{dosc}) and for $\omega_{\bf q}<\gamma/2$
one obtains

\begin{equation}
\frac{\Gamma_{\bf Q}}{T}\cong\pi[2\zeta+\Psi(1+\frac{\Gamma_{\bf Q}}{2\pi T})
-{\rm ln}\frac{\gamma}{2\pi T}]^{-1}, \label{zeta1}
\end{equation}
provided that $\gamma\gg 2\pi T$, and $\Psi(\cdot)$ is the digamma
function. Hence, in the regime $\Gamma_{\bf Q}\sim\alpha T$ and for moderate
$\alpha\!<2\pi$, we get $\alpha\sim\pi/(2\zeta)$ within leading order, which is
seen to agree very well with the exactly calculated values (inset to Fig.~1). 
\begin{figure}[b] 
\centering 
\epsfig{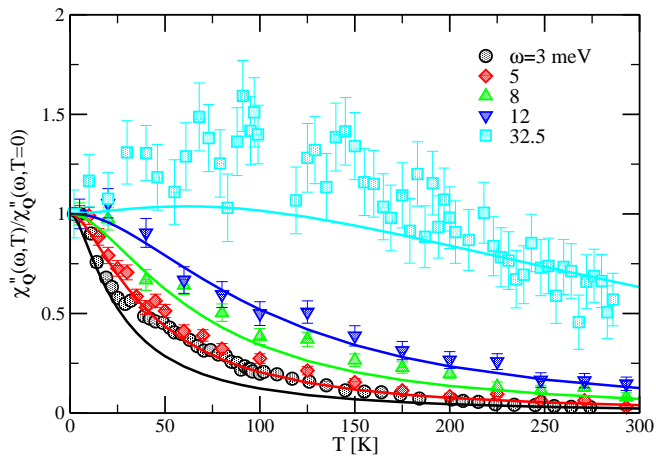}
\caption{Temperature evolution of $\chi''_{\mathbf Q}(\omega,T)$ for
UD YBCO with $x=0.45$ \cite{hink2} (symbols) compared with
theoretical result, Eq.~(\ref{chiim1}), with $\zeta=1.8$  and $\gamma$=40
meV (lines).} 
\label{fig2} 
\end{figure}

Obtained results can be directly put in relation with INS experiments
on UD cuprates. Several results have been recently reported for
$\chi''_{\bf q}(\omega)$, scanned in both ${\bf q}$ and $\omega$ for
different $T$. In particular, it has been observed that
$\chi''_{\bf Q}(\omega)$ can be well described by the simple
Lorentzian, Eq.~(\ref{chiim1}), and $\Gamma(T) \sim \alpha T$ has been
extracted. For UD YBCO with $x=0.35$ the authors \cite{stoc2} give
$\alpha=0.7$, whereas for LSCO weakly doped with $x=0.06$ Li scaling is well
described by $\alpha =0.18$ \cite{bao}. Electron-doped PLCCO
with doping $x=0.12$ also exhibits scaling \cite{wils2} with
$\alpha=0.44$. It is therefore plausible that different $\alpha$ can be within
our theory explained with different $\zeta$, Eq.(\ref{zeta}). 
Since $C_{\bf Q} \propto 1/\kappa^2_T$ the strongest dependence results from
$\kappa_T$ and experimental results on $\alpha
\propto 1/\zeta$ are consistent with the ranking
of measured (low-frequency) $\kappa$, i.e. $\kappa \sim 0.03$ (in
r.l.u.)  for YBCO \cite{stoc2}, $\kappa \sim 0.01$ for LSCO \cite{bao},
and $\kappa \sim 0.02$ for PLCCO \cite{wils1,wils2}. Since $\eta$ is a
robust quantity and rather well known from model calculations, only 
unknown $\gamma$ prevents so far more quantitative statements.

The above experimentally deduced $\alpha$ all require quite large $\zeta$,
hence one can expect, following Fig.~1, very low saturation $\Gamma^0_{\bf Q}$ for
$T \to 0$. On the other hand, mentioned INS data for YBCO and LSCO seem to
indicate quite substantial $\Gamma^0_{\bf Q}$. However, saturation of $\Gamma_{\bf
Q}$, setting in for $T<T_g$, is accompanied by the  simultaneous appearance
of the CP. This effect goes beyond here presented (simple) explanation
and requires within the PM the introduction of additional damping, as
discussed lateron.

Recent detailed INS measurements by Hinkov et al. on UD YBCO with
$x=0.45$ \cite{hink1,hink2} allow also for a more complete analysis of
$\chi''_{\bf Q}(\omega)$ as a function of $T$. Data
normalized at $T=0$, i.e., $\chi''_{\bf Q}(\omega,T)/\chi''_{\bf
Q}(\omega,T=0)$ are presented in Fig.~(\ref{fig2}) and compared with
theoretical ones, following from Eqs.(\ref{eqsum}),(\ref{dosc}), where the
only relevant parameter is $\zeta$.  
The overall agreement is quite satisfactory and is achieved for $\zeta\simeq
1.8$ implying $\Gamma^0_{\bf Q}\simeq 1.3$ meV. The marked dicrepancy between
theory and experimental data at $\omega=3\,$meV is tentatively associated
with a possible saturation of $\Gamma_{\bf Q}^0$ ($\chi''_{\bf
Q}(\omega,T=0)$) to a larger (smaller) value then predicted by theory.
Namely a better fit to experimental data based on $\zeta$ {\it and}
$\Gamma_{\bf q}^0$ as independent parameters, yields unchanged 
$\zeta$ but substantially larger $\Gamma_{\bf Q}^0\sim 4-5\,$meV,
which would correspond to a crossover temerature $T_g\sim\Gamma^0_{\bf
Q}/\alpha\sim 50$K. On the other hand, the poor agreement with the $32.5$
meV data should be attributed to the breakdown of $\omega/T$ scaling since
$\omega>\gamma/2$. 

\begin{figure}[htb] 
\centering 
\epsfig{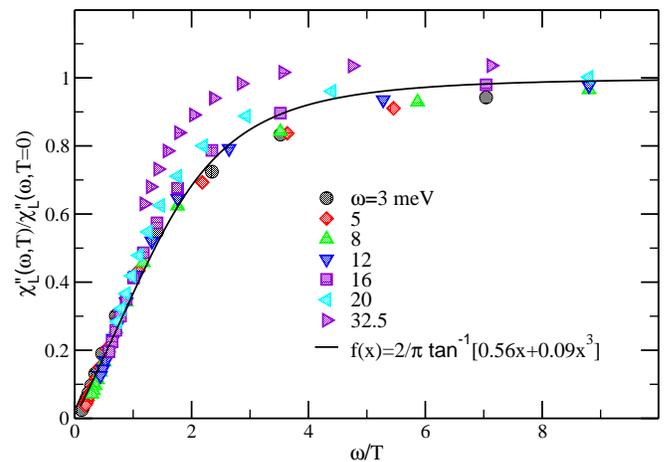}
\caption{Scaling of theoretical  normalised $\chi''_L(\omega,T)$ based on
$\zeta$ and $\gamma$ as in Fig.~(\ref{fig2}). For comparison, the scaling
function $f(x)$ as frequently used in fits to experiments is also plotted.}
\label{fig3} 
\end{figure}

Experimental results for {\bf q}-integrated $\chi''_L(\omega)$
in the same regime also reveal the $\omega/T$ scaling and have been
several times fitted with the ansatz involving ${\rm tan}^{-1}(x)$, as
already noted in the introduction. However, the fit did not find so far a deeper
theoretical background. It has been nevertheless shown that results of
the present theory are quite close to this description \cite{prel}. In fact,
assuming a Lorentzian dependence of $\chi_{\bf q}^0$ on $\bf q$ and
invoking the relation $\Gamma_{\bf q} =\eta/(\gamma\chi_{\bf q}^0)$ a simple
calculation yields $\rm tan^{-1}(A_1\omega/T)$ with $A_1\sim 1/\alpha$,
provided that  $\kappa\ll1$ but $\kappa^2\chi^0_{\bf Q}\sim$ const. Note,
however, that quite generally $\chi_{\bf q}^0$ obtained from
Eq.~(\ref{eqsum}) decays at least as a double-Lorentzian, so that
$\chi_L''(\omega,T)$ is a more complicated function, allowing even for
nonanalytic behavior in $\omega$ for $\zeta\gg1$.

In Fig.~(3) we next present the theoretically obtained data for the scaling function
$f(\omega/T)=\chi''_{\rm L}(\omega,T)/\chi''_{\rm L}(\omega, T=0)$ for several
different $\omega$ and the temperature range $0-300$ K based on the same set
of data for $\gamma$ and $\zeta$ as discussed in 
relation to Fig.~(2). While the $\omega/T$ scaling is quite remarkable it
breaks down at higher energies, as expected. We also plot the scaling
function $f(x)=2/\pi{\rm tan}^{-1}(A_1x+A_2x^3)$ often employed in fits to
experimental data: the `best-fit' coefficients $A_1\sim 0.56$ and $A_2\sim
0.09$ are in reasonable agreement with values $A_1\sim 0.49,A_2\sim 0.12$
determined directly from experimental data \cite{hink2}.

{\it CP response:} The advantage of the memory-function
formalism is that the emergence of the CP at $T<T_g$ in the
spin response can be as well treated within the same framework.  One
has to assume that unlike in a PM the mode damping
$\gamma_{\bf q}$ is not constant but may acquire an additional low frequncy
contribution.  In particular we can take
\begin{equation}
\tilde M_{\bf q} \sim i \gamma - \delta^2/(\omega+i\lambda),
\label{cent}
\end{equation}
with $T$-dependent $\delta$ and $\lambda$, which via Eq.(\ref{dosc}) leads
to  $\chi''_{\bf q}(\omega)$ of the form used also to analyse experimental
INS data  for YBCO with $x=0.35$ \cite{stoc1}. It has been first employed
for the analysis of the CP in ferroelectrics \cite{halp}, 
assuming a coupling of a damped mode to a slowly fluctuating object. 

For $\lambda\to0$ but $\delta^2/\lambda\gg \gamma$ the modified $\tilde
M_{\bf q}$ leads to two distinct energy scales and hence to two contributions
to spin dynamics, i.e., the CP part $\chi^c_{\bf q}(\omega)$ and
the regular contribution $\chi^r_{\bf q}(\omega)$,
\begin{equation}
\chi^c_{\bf q}(\omega)\sim \frac{\chi_{\bf
q}^0\Gamma_c}{\Gamma_c-i\omega},\,\, \chi^{r}_{\bf
q}(\omega)\sim \frac{\chi_{\bf q}^{r0}\Gamma_r}{\Gamma_r-i\omega},
\,\,{\bf q}\sim {\bf Q},
\label{cpeq}
\end{equation}
valid for $\omega<\lambda$ and $\lambda<\omega\ll\gamma$, 
respectively. Thus, below $T_g$ the new scales are set by
$\Gamma_r=\Omega_{\bf q}^2/\gamma$
and $\Gamma_c=(\eta/\delta^2)\lambda/\chi_{\bf q}^0$, with $\chi_{\bf
q}^{r0}=\eta/\Omega_{\bf q}^2$ and $\Omega_{\bf q}^2=\omega_{\bf
q}^2+\delta^2$. If one assumes that $\delta$ saturates at low $T$, as is
manifest by saturation of $\Gamma_r$ \cite{stoc2},
$\Gamma_c\!\propto\!\lambda/\chi_{\bf q}^0$ becomes
the smallest energy scale, resulting in a quasielastic peak which is,
according to Eq.~(\ref{cpeq}), a Lorentzian of width $\Gamma_c$, with the
static value $\chi_{\bf q}(\omega\to\!0)=\chi^0_{\bf q}$,  as required by 
Eq.~(\ref{chiq}). However, the real part drops for $\omega>\lambda$ to
the value $\chi^{r0}_{\bf q}$, hence the CP contribution is  
$\Delta \chi_{\bf q}=\chi_{\bf q}^0(\delta/\Omega_{\bf q})^2\approx\chi_{\bf q}^0$.  

It should be pointed out that within such a formalism a modified
$\tilde M_{\bf q}$ does not (necessarily) induce a change of
equal-time correlations as $C_{\bf q}$, in particular the characteristic
low-$\omega$ width $\kappa$, remains unchanged, a fact observed
experimentally \cite{stoc1,stoc2}. Moreover, as the CP is too narrow
to affect the moments of $\chi^{\prime\prime}_{\bf q}(\omega)$, $\eta_{\bf q}$ is determined
solely by the regular part. It is then plausible to assume
that also $\chi^{r\prime\prime}_{\bf q}(\omega)$ remains constant on
lowering $T<T_g$, as also observed experimentally in Li-doped LSCO \cite{bao}
where below $T_g \sim 50$~K the ${\bf q}$-integrated regular part
takes the form $\chi''_L(\omega) \propto \omega/(\omega^2+ \Gamma^2)$ with
$T$-independent $\Gamma$. The CP contribution then follows from the sum rule  
\begin{equation}
T\Delta\chi_{\bf q}=C_{\bf q}^c(T)=C_{\bf q}-C^{r}_{\bf q}(T), 
\label{sumc}
\end{equation}
with the regular part  given by \cite{nobe} 
\begin{equation}
C^r_{\bf q}(T)=\frac{\eta}{\pi\gamma}{\rm ln}\frac{\gamma}{\Gamma_r}+\frac{\pi
T}{\gamma\Gamma_r},\quad T\ll T_g,
\end{equation}
again consistent with experiment on $x=0.35$ YBCO \cite{stoc2} where below
$T\sim 50$~K the (integrated) intensity transfers from a broad inealstic scattering to a 
CP while conserving the (total) sum rule. 

From the above considerations it follows that the onset of CP can
be explained as a dynamical freezing phenomenon. However, it is
evident that the appearance of the anomalous (singular at low
$\omega$) damping $\tilde M_{\bf q}(\omega)$ is not
consistent with normal metallic particle-hole excitations. We can
speculate that to obtain such a form one has to invoke an inhomogeneous
(disordered) system with predominantly localized charge carriers preventing
low-$\omega$ scattering, resulting in a nearly diverging $\tilde M_{\bf
q}(\omega\to\! 0)$. In any case the transition to CP-type dynamics
can be associated with a significant softening of $\lambda(T<T_g)$.  

In conclusion, we have shown that the approach presented gives a
consistent explanation of the $\omega/T$ scaling both in $\chi''_{\bf
q}(\omega)$ as well as in $\chi_L''(\omega)$. It is  based on two well established
experimental facts: the overdamped nature of the response and the saturation
of effective inverse correlation length $\kappa(\omega)$ at low $\omega$ and
$T$ leading to the anomalous slowing down, i.e., gradual freezing of the
dynamic spin response, a phenomenon of entirely quantum origin. The theory
can as well account for the INS results in UD cuprates, yielding an
explanation for the nonuniversal parameter $\alpha=\Gamma_{\bf Q}/T$.  

The appearance of the CP for $T<T_g$ is easily incorporated into the
formalism via the (almost) singular $\omega$- and $T$-dependent damping
${\tilde M}_{\bf q}(\omega)$. However, the question as to the nature of CP,
whether of intrinsic origin, based on spin-glass-like arguments, as
suggested, e.g., in \cite{stoc1}, or of some other, possibly even extrinsic
origin, remains to be settled.  

The authors would like to thank Dr. V. Hinkov for using his data prior to
publication and for fruitful discussion.
This work was supported by Slovenian Research Agency (ARRS) under grant
No. P1-0044-106.


\begin{thebibliography}{99}
\bibitem{keim} B.\ Keimer {\it et al.}, Phys.\ Rev.\ Lett. \textbf{67},
1930 (1991); Phys.\ Rev.\ B \textbf{46}, 14034 (1992).
\bibitem{kast} for a review see M.\ A.\ Kastner, R.\ J.\ Birgeneau, 
G.\ Shirane, and Y.\ Endoh, Rev.\ Mod.\ Phys.\ \textbf{70}, 897
(1998).
\bibitem{ster} B.\ J.\ Sternlieb, G.\ Shirane, J.\ M.\ Tranquada, M.\ Sato,
and S.\ Shamoto, Phys.\ Rev.\ B \textbf{47}, 5320 (1993).
\bibitem{kaku} K.\ Kakurai, S.\ Shamoto, T.\ Kiyokura, M.\ Sato ,J.\ M.\
Tranquada, and G.\ Shirane, Phys.\ Rev.\ B \textbf{48},
3485 (1993).
\bibitem{bao} W.\ Bao, Y.\ Chen, Y.\ Qiu, and J. L.\ Sarrao, 
Phys.\ Rev.\ Lett. \textbf{91}, 127005 (2003).
\bibitem{stoc1} C.\ Stock, W.\ J. L.\ Buyers, Z.\ Yamani, C.\ L.\ Broholm, 
J.-H.\ Chung, Z.\ Tun, R.\ Liang, D.\ Bonn, W.\ N.\ Hardy, and R.\ J.\
Birgeneau, Phys.\ Rev.\ B \textbf{73}, 100504(R) (2006).
\bibitem{stoc2} C.\ Stock, W.\ J. L.\ Buyers, Z.\ Yamani, Z.\ Tun, 
R.\ J.\ Birgeneau, R.\ Liang, D.\ Bonn, and W.\ N.\ Hardy, 
Phys.\ Rev.\ B \textbf{77}, 104513 (2008).
\bibitem{hink1} V.\ Hinkov, D.\ Haug, B.\ Fauque, P.\ Bourges, Y.\ Sidis, 
A.\ Ivanov, C. Bernhard, C.\ T.\ Lin, and B.\ Keimer, Science
\textbf{319}, 597 (2008).
\bibitem{hink2} V.\ Hinkov, Ph.\ D.\ thesis (2007), unpublished.
\bibitem{wils1} S.\ D.\ Wilson, S.\ Li, H.\ Woo, P.\ Dai, H.\ A.\ Mook, 
C. D.\ Frost, S.\ Komiya, and Y.\ Ando, Phys.\ Rev.\
Lett. \textbf{96}, 157001 (2006).
\bibitem{wils2} S.\ D.\ Wilson, S.\ Li, P.\ Dai, W.\ Bao, J.-H.\ Chung, 
H.\ J.\ Kang, S.-H.\ Lee, S.\ Komiya, Y.\ Ando, and Q.\ Si,  Phys.\ Rev. B 
\textbf{74}, 144514 (2006).
\bibitem{varm} C.\ M.\ Varma, P.\ B.\ Littlewood, S.\
Schmitt-Rink, E.\ Abrahams, and A.\ E.\ Ruckenstein, Phys.\ Rev.\ 
Lett. \textbf{63}, 1996 (1989).
\bibitem{chak} S.\ Chakravarty, B.\ I.\ Halperin, and D.\ R.\ Nelson,
Phys.\ Rev.\ Lett. \textbf{60}, 1057 (1988); Phys.\ Rev.\ B \textbf{39},
2344 (1989).  
\bibitem{aepp} G.\ Aeppli, T.\ E.\ Mason, S.\ M.\
Hayden, H.\ A.\ Mook, and J. Kulda, Science, \textbf{278}, 1432 (1997).
\bibitem{schr} A.\ Schroder, G.\ Aeppli, E.\ Bucher, R. Ramazashvili, 
and P.\ Coleman, Phys.\ Rev.\ Lett. \textbf{80}, 5623 (1998).
\bibitem{prel}  P.\ Prelov\v sek, I.\ Sega, and J.\ Bon\v ca, 
Phys.\ Rev.\ Lett. \textbf{92}, 027002 (2004).
\bibitem{sega} I.\ Sega, P.\ Prelov\v sek, and J.\ Bon\v ca, 
Phys.\ Rev.\ B \textbf{68}, 054524 (2003).
\bibitem{halp} B.\ I.\ Halperin, and C.\ M.\ Varma, 
Phys.\ Rev.\ B \textbf{14}, 4030 (1976); S.\ M.\ Shapiro, J.\ D. Axe, G.\ Shirane, and T.\ Riste,
Phys.\ Rev.\ B \textbf{6}, 4332 (1972); I.\ Sega, Ph.\ D.\ thesis (1979), unpublished.
\bibitem{nobe}Note that this estimate follows from Eq.~(\ref{zeta1}) on assuming that
$\Gamma_{\bf q}(T)$ saturates to (almost) $T$-independent $\Gamma_r$ below $T_g$.
\end{thebibliography}
\end{document}